\documentclass[useAMS,usenatbib,usegraphicx]{mn2e}
\newcommand{\der}{\mathrm{d}}
\newcommand{\reff}{r_\mathrm{eff}}
\newcommand{\piz}{\Pi_{zz}}
\newcommand{\pir}{\Pi_{RR}}
\newcommand{\pip}{\Pi_{\phi \phi}}
\newcommand{\pix}{\Pi_{xx}}
\newcommand{\piy}{\Pi_{yy}}

\newcommand{\vcirc}{v_\mathrm{circ}}
\newcommand{\fmax}{f_S}
\newcommand{\frad}{f_\mathrm{rad}}
\newcommand{\felz}{f_{2I}}
\newcommand{\shapefunc}{f^+}

\newcommand{\sau}{SAURON }
\newcommand{\coma}{COMA }

\title[]{The flattening and the orbital structure of early-type galaxies and collisionless $N$-body binary disk mergers}
\author[J. Thomas et al.]{J. Thomas$^{1,2}$\thanks{E-mail: jthomas@mpe.mpg.de}, R. Jesseit$^{1}$, R. P. Saglia$^{1,2}$,  
R. Bender$^{1,2}$, A. Burkert$^{1}$, \newauthor E.~M. Corsini$^{3}$, K. Gebhardt$^{4}$, 
J. Magorrian$^{5}$, T. Naab$^{1}$, \newauthor D. Thomas$^{6}$ and G. Wegner$^{7}$\\
$^{1}$Universit\"atssternwarte M\"unchen, Scheinerstra\ss e 1, D-81679 M\"unchen, Germany\\
$^{2}$Max-Planck-Institut f\"ur Extraterrestrische Physik, Giessenbachstra\ss e, D-85748 Garching, Germany\\
$^{3}$Dipartimento di Astronomia, Universit\`a di Padova, vicolo dell'Osservatorio 3, I-35122 Padova, Italy\\
$^{4}$Department of Astronomy, University of Texas at Austin, C1400, Austin, TX78712, USA\\
$^{5}$Theoretical Physics, Department of Physics, University of Oxford, 1 Keble Road, Oxford U.K., OX1 3NP\\
$^{6}$Institute of Cosmology and Gravitation, Mercantile House, University of Portsmouth, Portsmouth, PO1 2EG, UK\\
$^{7}$Department of Physics and Astronomy, 6127 Wilder Laboratory, Dartmouth College, Hanover, NH 03755-3528, USA}
\begin{document}

\date{Submitted to MNRAS ------; Accepted ------------}

\pagerange{\pageref{firstpage}--\pageref{lastpage}} \pubyear{2005}

\maketitle

\label{firstpage}
\begin{abstract}
We use oblate axisymmetric dynamical models including dark halos to determine the orbital structure 
of intermediate mass to massive Coma early-type
galaxies. We find a large variety of orbital compositions. 
Averaged over all sample galaxies the unordered stellar kinetic energy
in the azimuthal and the radial direction are of the same order, but they can differ by up to
40 percent in individual systems. 
In contrast, both for rotating and non-rotating galaxies the vertical kinetic 
energy is on average smaller than in the other two directions. This implies that 
even most of the rotating ellipticals
are flattened by an anisotropy in the stellar velocity 
dispersions. Using three-integral axisymmetric toy models we show that flattening by 
stellar anisotropy maximises the entropy for a given density distribution. 
Collisionless disk merger remnants are radially 
anisotropic. The apparent lack of strong radial anisotropy in
observed early-type galaxies implies that they may not
have formed from mergers of disks unless the influence of dissipational processes was
significant.
\end{abstract}


\section{Introduction}
\label{sec:intro}
The way in which a galaxy has assembled its stars is reflected in the distribution of stellar 
orbits. For example, collisionless
$N$-body collapse simulations predict a predominance of radial orbits in the final
remnant \citep{vanAl82}. In contrast, collisionless galaxy merger simulations predict a variety of 
orbital compositions, depending on progenitor 
properties \citep{Bar92,Her92,Her93}, the merging geometry \citep{Wei96,Dub98}, 
the progenitor mass ratios \citep{Nab03,Jes05} and the presence of
dissipational components \citep{Nab06,Jes07}. 
Stars in galaxies are approximately collisionless and the orbital structure -- once a
galaxy has approached a quasi-steady-state -- is conserved for a long time. 
To a certain extent then, the
assembly mechanism of early-type galaxies can be constrained from their present-day
orbital structure.

A global characteristic of the distribution of stellar orbits is its
anisotropy.
Traditionally, anisotropies of elliptical galaxies have been inferred from 
the ($v/\sigma,\epsilon$) diagram. In particular, the rotation of bright ellipticals has been
shown to be insufficient to account for their flattening \citep{Bin78}. However, whether fainter, fast-rotating
ellipticals are flattened by rotation is less easy to determine from 
the ($v/\sigma,\epsilon$) diagram, because isotropic as well as anisotropic systems can
rotate. In fact, fully general axisymmetric dynamical models recently have revealed 
an anisotropic orbital
structure in even the flattest, fast rotating objects \citep{Cap07}. One goal of this paper is to 
investigate numerically the connection between anisotropy, rotation 
and flattening in spheroidal stellar systems.

In addition, we present global anisotropies for a sample of Coma early-type galaxies.
These anisotropies are derived by analysing long-slit stellar absorption
line kinematics with axisymmetric orbit models. Our dynamical models include 
dark matter halos.
Previous anisotropy determinations for larger samples of ellipticals (including
dark matter) were restricted to round and non-rotating systems, assuming spherical 
symmetry \citep{G01,Mag01}. 
Spherical models do not account for galaxy flattening. In the simplest case, a flattened
system is axially symmetric. Early axisymmetric
models, however, did not cover all possible anisotropies 
(and orbital structures, respectively; e.g. \citealt{Car95}). Fully general, orbit-based
axisymmetric dynamical models have so far only been applied to the inner regions of
ellipticals and the orbital analysis was made under the assumption that mass follows 
light (e.g. \citealt{Geb03} and \citealt{Cap06}). By the mass-anisotropy degeneracy, 
the neglect of dark matter could translate to a systematic bias in the corresponding
orbital structure (e.g. \citealt{For08}). Comparison of anisotropies derived with and
without dark matter will allow one to quantify such a possible bias.

We also discuss anisotropies derived from modelling mock observations of
synthetic $N$-body merger remnants. One motivation to do so is that dynamical models 
of axisymmetric systems may not be unique. For example, the deprojection of an
axisymmetric galaxy is intrinsically degenerate \citep{Ryb87}. Uncertainties in the intrinsic shape thereby
propagate into uncertainties on the derived masses and anisotropies (e.g. \citealt{Tho07b}).
Moreover, the reconstruction of an axisymmetric orbital system 
is suspected to be further degenerate with the recovered mass (e.g. the discussion 
in \citealt{Val04}). The case for a generic degeneracy, beyond the effects of
noise and incompleteness of the data, is still uncertain (e.g. \citealt{Mag06}). Numerical 
studies of a few idealised axisymmetric toy models indicate degeneracies to be moderate when 
modelling realistically noisy data sets
(\citealt{Kra05}, \citealt{Tho05}). Since we know the true structure
of our $N$-body modelling targets, we can extend on these studies and further investigate 
potential systematics in the models over a broader sample of test cases. 

Another motivation to model $N$-body merger remnants is to probe whether
ellipticals have formed by merging. This requires a comparison
of the orbital structure in real ellipticals with predictions of $N$-body 
simulations (e.g. \citealt{Bur05,Bur07}). However, because of the symmetry assumptions 
in models of real galaxies, it is not straight forward to compare {\it intrinsic} properties 
of $N$-body simulations with {\it models} of real galaxies. To avoid the related
systematics, we here compare models of real galaxies with {\it similar models} of 
synthetic $N$-body merger simulations and both are indicative for true 
differences between real galaxies and merger predictions.

The galaxy and $N$-body merger samples and the modelling technique are briefly outlined
in Sec.~\ref{sec:data}. Toy models of various flattening and anisotropy are discussed
in Sec.~\ref{sec:theory}. The anisotropies of real galaxies are presented in 
Sec.~\ref{sec:galaxies} and compared with models of $N$-body merger remnants
in Sec.~\ref{sec:merger}. Implications for the formation process of early-type galaxies
are discussed in Sec.~\ref{sec:discussion} and we summarise our results 
in Sec.~\ref{sec:summary}. The influence of regularisation and the inclusion of dark
matter halos on reconstructed galaxy anisotropies is discussed in App.~\ref{sec:app}.
In App.~\ref{app:entropy} we briefly discuss the connection between anisotropy 
and the shape of the circular velocity curve in maximum
entropy models.
We assume that the Coma cluster is at a distance of 100 Mpc.

\begin{table}\centering
\begin{tabular}{lccc||cccc}
\multicolumn{1}{c}{GMP} & NGC & $\epsilon$ & $\delta$ & $\beta$ & $\gamma$ & $(v/\sigma)^*$\\
\multicolumn{1}{c}{(1)} & (2) & (3) & (4) & (5) & (6) & (7)\\
\hline
0144 & 4957    &  $0.48$ & $0.44$  & $0.36$  & $-0.27$ &  $0.42$\\
0282 & 4952    &  $0.57$ & $0.39$  & $ 0.45$ & $0.17$  &  $0.64$\\
0756 & 4944    &  $0.61$ & $0.22$  & $ 0.29$ & $0.10$  &  $0.83$\\
1176 & 4931    &  $0.65$ & $0.22$  & $ 0.22$ & $-0.16$ &  $0.91$\\
1750 & 4926    &  $0.17$ & $0.21$  & $ 0.26$ & $0.11$  &  $0.17$\\
1990 & IC 843  &  $0.54$ & $-0.03$ & $ 0.13$ & $0.16$  &  $1.11$\\
2417 & 4908    &  $0.28$ & $0.14$  & $ 0.24$ & $0.21$  &  $1.02$\\
2440 & IC 4045 &  $0.58$ & $-0.08$ & $ 0.06$ & $0.07$  &  $1.51$\\
2921 & 4889    &  $0.35$ & $0.19$  & $ 0.31$ & $0.30$  &  $0.09$\\
3329 & 4874    &  $0.11$ & $0.05$  & $-0.12$ & $-0.36$ &  $0.63$\\
3414 & 4871    &  $0.42$ & $0.16$  & $0.18$  & $0.04$  &  $0.97$\\ 
3510 & 4869    &  $0.17$ & $0.13$  & $ 0.13$ & $-0.01$ &  $0.71$\\
3792 & 4860    &  $0.28$ & $0.17$  & $ 0.13$ & $-0.08$ &  $0.29$\\
3958 & IC 3947 &  $0.36$ & $0.20$  & $ 0.12$ & $-0.21$ &  $1.03$\\
4822 & 4841A   &  $0.22$ & $0.12$  & $ 0.20$ & $0.17$ &  $0.41$\\
4928 & 4839    &  $0.34$ & $0.30$  & $ 0.31$ & $0.02$  &  $0.39$\\
5279 & 4827    &  $0.23$ & $0.15$  & $ 0.16$ & $0.02$  &  $0.39$\\
5568 & 4816    &  $0.22$ & $0.21$  & $ 0.36$ & $0.37$  &  $0.27$\\
5975 & 4807    &  $0.21$ & $0.06$  & $ 0.05$ & $-0.03$ &  $1.29$\\
\end{tabular}
\caption{Summary of \coma galaxy anisotropies. (1-2): galaxy id (GMP numbers from \citealt{GMP}); 
(3): intrinsic ellipticity $\epsilon$; 
(4-6): anisotropy parameters $\delta$, $\beta$
and $\gamma$ (cf. equations \ref{eq:delta}-\ref{eq:gamma}) 
of the best-fit dynamical model; (7): $(v/\sigma)^*$, i.e. $(v_\mathrm{max}/\sigma_o)$
normalised by the approximate value $\sqrt{\epsilon_\mathrm{obs}/(1-\epsilon_\mathrm{obs})}$ of an (edge-on) 
isotropic rotator with the same flattening. 
Note that $(v/\sigma)^*$ is an observable, i.e. it combines {\it observed}
ellipticities $\epsilon_\mathrm{obs}$ (from column 10 of Tab.~1 of \citealt{Meh00}) and observed velocities $v_\mathrm{max}$ 
and $\sigma_o$, without reference to any dynamical model.}
\label{tab:aniso}
\end{table}

\section{Data and basic definitions}
\label{sec:data}
A complete description of a stellar system is given by its distribution
function $f$ (DF; the density in 6-dimensional phase-space). 
In a steady-state system the DF $f$ depends on the phase-space coordinates only through
the integrals of motion \citep{Lyn62}. Axisymmetric 
potentials, which are considered here, admit the two classical integrals of
motion energy ($E$) and z-component of the angular momentum ($L_z$). In addition, many
orbits in astrophysically relevant potentials are characterised by another, non-classical, 
so-called third integral ($I_3$; \citealt{Con63}). Since integrals of motion label orbits and vice-versa,
a steady-state system can be viewed as a
superposition of orbits, each with constant phase-space density. Let $f(i)$ denote 
the phase-space density along orbit $i$, then the total amount of light $w(i)$ on the orbit 
equals $w(i) = f(i) \times V(i)$ ($V(i)$ is the orbit's phase-space volume). The DF -- or the weights
of a suitable orbit superposition model -- determine the
spatial density $\rho$ and intrinsic velocity dispersions $\sigma$ via
\begin{equation}
\label{dens}
\rho = \int f \, \der^3 v
\end{equation}
and
\begin{equation}
\label{e:sig}
\sigma_{ij}^2 = \frac{1}{\rho} \, \int f \, (v_i-\overline{v_i})(v_j-\overline{v_j}) \, \der^3 v,
\end{equation}
\begin{equation}
\label{e:vav}
\overline{v_i} = \frac{1}{\rho} \, \int f \, v_i \, \der^3 v.
\end{equation}
In the following we will only consider $i,j \in \{z,x,\phi,R\}$,
where $z$ is the short axis of the density distribution, $\phi$ is the azimuth around this axis,
$x$ is a fixed Cartesian coordinate parallel to the equatorial plane and $R$ is a 
cylindrical radius. Let
\begin{equation}
\label{pidef}
\Pi_{ii}=\int \rho \sigma_{ii}^2 \, \der^3 r
\end{equation}
denote the total\footnote{In the following we will only consider anisotropies
(and kinetic energies, respectively) inside the effective radius, 
because this is the radius inside which kinematical
observations are typically available to constrain the orbital structure of real galaxies.} 
(unordered) kinetic energy
in coordinate direction $i$, then the global anisotropy of an axisymmetric stellar system 
can be quantified, for example, by the ratios
\begin{equation}
\label{eq:delta}
\delta \equiv 1 - \frac{\piz}{\pix}
\end{equation}
\begin{equation}
\label{eq:beta}
\beta \equiv 1 - \frac{\piz}{\pir}
\end{equation}
and
\begin{equation}
\label{eq:gamma}
\gamma \equiv 1 - \frac{\pip}{\pir}
\end{equation}
\citep{Cap07}. In axisymmetric systems, the three anisotropy parameters are related via
\begin{equation}
\delta = \frac{2\beta-\gamma}{2-\gamma}.
\end{equation}
Non-rotating, isotropic spherical systems as well as 
classical isotropic rotators obey $\delta = \beta = \gamma = 0$. 

The DF of real galaxies is not known, but has to be reconstructed from photometric
and kinematic observations. In the next two subsections we will describe the
two samples of real and simulated galaxies discussed in this paper and will briefly outline 
our modelling method.

\subsection{Coma early-types}
Our sample of observed galaxies (\coma in the following) consists of 19 Coma early-types 
from \citet{Tho07,Tho08}. It comprises 
2 central cD galaxies,
10 ordinary giant ellipticals and 7 S0 or intermediate galaxies with luminosities 
$-20.3 \ge M_B \ge -22.56$ (a single fainter object with $M_B=-18.8$
is also included in the sample). The galaxies are drawn from the luminosity limited 
sample of \citet{Meh00} and are distributed all over the cluster. High-resolution 
radial profiles of surface brightness, 
ellipticity and isophotal shape parameters $a_4$ and $a_6$ (up to $a_{12}$ in some cases;
cf. \citealt{Ben87} for a definition of the isophotal shape parameters)
derived from a combination of HST and ground-based imaging were
used to calculate the deprojected 3d luminosity distribution for several inclinations. 
The photometric data are complemented by long-slit stellar 
absorption line kinematics along $2-4$ position-angles per galaxy. The kinematic
data consists of 
radial profiles of mean velocity, velocity dispersion and higher-order
moments of the line-of-sight velocity distribution and reach out to
$1-4 \, \reff$. Details about the photometric and kinematic data can be found in
\citet{Jor94}, \citet{Meh00}, \citet{Weg02}, \citet{Cor08} and \citet{Tho08}.

These data were modelled with our 
implementation of Schwarzschild's (1979) orbit superposition 
technique for axisymmetric potentials \citep{Ric88,Geb00,Geb03,Tho04}. 
For each galaxy, we probed for a variety of mass models, composed of a stellar mass
density (from the deprojected light profile) and a parametric dark halo profile.
The parameter space for the mass models spans the inclination,
the stellar mass-to-light ratio and the dark halo parameters.
In each trial potential the best-fit orbit model is calculated by maximising
\begin{equation}
\label{orbitmaxs}
S - \alpha \chi^2,
\end{equation}
where $\chi^2$ quantifies deviations between observed and modelled kinematics\footnote{We
use the full information of (binned) line-of-sight velocity distributions when
fitting real galaxies and $N$-body merger remnants.}.
The function 
\begin{equation}
\label{sgen}
S = - \sum w(i) \ln \frac{w(i)}{\Omega(i)}
\end{equation}
is used to smooth the orbit models. 
In the absence of any other constraints
the maximisation of S yields orbital weights $w(i) \propto \Omega(i)$ \citep{Ric88},
such that the yet not specified $\Omega(i)$ can be regarded as weight-factors for
the $w(i)$. When modelling real galaxies or mock 
observations of $N$-body merger remnants, we assume that there is no preferred
region is phase-space and each orbit is given an a priori-weight equal to
its phase-space volume: $\Omega(i) \equiv V(i)$. Then, 
\begin{equation}
\label{maxs}
S = - \sum w(i) \ln \frac{w(i)}{V(i)} \approx - \int f \ln f \, \der^3 r \, \der^3 v 
\end{equation}
equals the Boltzmann entropy, which drives models towards a 
constant density in phase-space.

The (binned) deprojected luminosity density is used as 
a boundary condition to solve equation (\ref{orbitmaxs})
and the regularisation parameter $\alpha$ in equation (\ref{orbitmaxs}) has been calibrated 
by means of 
Monte-Carlo simulations \citep{Tho05}. The final, best-fit orbit model is obtained from
a $\chi^2$-analysis.

\subsection{$N$-body merger remnants}
We have applied the same modelling code to mock observations of 
synthetic $N$-body merger simulations. In brief, we have modelled 
six merger remnants, each projected along its three principal axes (models
of projections along the long, intermediate and short axis of the merger
remnants will be shortly referred to as X, Y and Z-models later on). The six 
merger remnants are taken from the sample of collisionless disk+bulge+halo mergers
of \citet{Nab03}. They have mass ratios between 1:1 and 4:1 and sample
the entire distribution of intrinsic shapes and orbital make-ups, 
including extreme cases. An orbital analysis of the $N$-body systems is given in 
\citet{Jes05}. We have simulated typical Coma observations for each projection: 
the merger remnants were placed at a distance of 100 Mpc and
photometric and kinematic profiles with similar resolution and spatial
coverage as in the \coma sample have been extracted.
For a detailed discussion of the models the reader is referred to \citet{Tho07b}.

%
\begin{figure*}\centering
\begin{minipage}{170mm}
\centering
\includegraphics[width=165mm,angle=0]{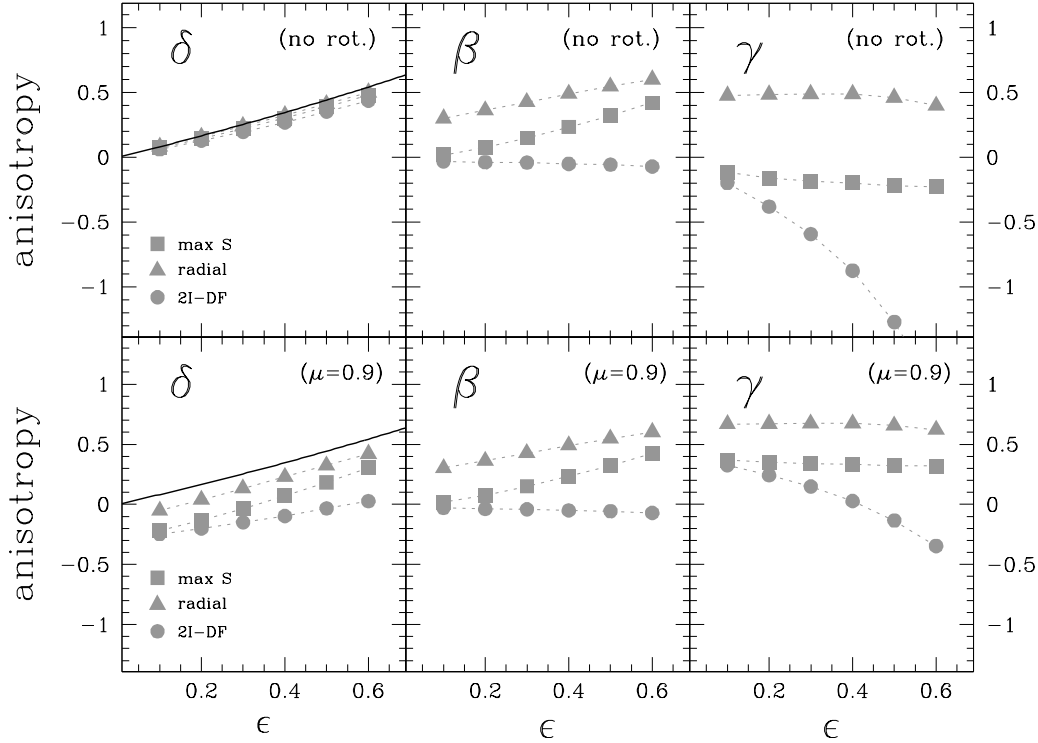}
\caption{Anisotropy parameters $\delta$, $\beta$ and $\gamma$ (as labelled in the
panels) versus intrinsic ellipticity $\epsilon$ for the toy models $\fmax$ (squares), 
$\felz$ (circles) and $\frad$ (triangles). Toy models of the same type are 
connected by dotted lines. Top: models without rotation; bottom: 
models with rotation ($\mu = 0.9$; cf. equation \ref{flammu}); light symbols:
without dark matter halo; solid lines: tensor 
virial theorem applied to oblate spheroids with constant flattening.}
\label{aniso_prof}
\end{minipage}
\end{figure*}

\section{Toy models}
\label{sec:theory}
Oblate stellar systems can owe their shapes to a variety of different orbital
configurations. Classically, one has often distinguished between two proto-typical 
cases:
flattening by rotation and flattening by anisotropy. Thereby, flattening by rotation
is used to term an otherwise round and isotropic system which appears flattened -- and
rotating -- by extra-light on near-equatorial, high angular-momentum orbits (populated with the same
sense of rotation). Flattening by anisotropy refers to systems with a depression 
of stars with high velocities perpendicular to the equatorial plane ($\piz < \pir = \pip$).
However, in fact there are infinitely many orbit superpositions that account
for a given galaxy shape. Some of these are discussed in
\citet{DG93}. Different orbital structures can be distinguished by
their different anisotropies. In the following we will numerically construct 
(self-consistent) toy models that are designed to (1) reproduce a given, flattened, 
density distribution exactly, but (2) have different intrinsic anisotropies.

\subsection{Self-consistent models without rotation}
\label{subsec:toy}
The models are orbit-based and similar to those 
described in Sec.~\ref{sec:data}. However, here we only require 
the models to reproduce a given density 
($\alpha = 0$ in equation \ref{orbitmaxs}). Various expressions for the
factors $\Omega(i)$ in equation (\ref{sgen}) will be used to impose different
anisotropy structures (see below).

For our simple toy models we assume a stellar density
\begin{equation}
\label{eq:dehnen}
\rho \propto m^{-1} (m+1)^{-3}
\end{equation}
\citep{Her90} with
\begin{equation}
\label{eq:flattening}
m^2 = R^2 + \frac{z^2}{q^2}.
\end{equation}
Equations (\ref{eq:dehnen}) and (\ref{eq:flattening}) describe systems with 
constant flattening $q$. They approximate the light profiles of elliptical galaxies
reasonably well.

\subparagraph*{Flattening and maximum entropy.}
Let $\fmax$ denote the DF
that maximises the entropy of equation (\ref{maxs}) subject to the
density constraints. The 
squares in the top panel of Fig.~\ref{aniso_prof} illustrate the connection between 
anisotropy and flattening for $\fmax$: the three panels show 
the anisotropy parameters from equations (\ref{eq:delta}-\ref{eq:gamma}) 
as a function of the intrinsic ellipticity $\epsilon \equiv 1-q$ 
(cf. equation \ref{eq:flattening}).
While $\delta$ and $\beta$ increase with flattening, $\gamma$ 
is roughly constant. In maximum-entropy models the flattening thus arises
from a suppression of energy in $z$-direction, while the balance between the
energies in $R$ and $\phi$ is roughly conserved. In this sense, the maximum entropy
models $\fmax$ resemble the classical case of flattening by anisotropy. The
only difference is that $\gamma \ne 0$ (cf. App.~\ref{app:entropy} for a discussion
of $\gamma$). Note that we calculated the toy models with the same library setup as
used for the Coma galaxy models.

\subparagraph*{Flattening by a classical $f=f(E,L_z)$.}
A classical two-integral 
DF $\felz$, which only depends on $E$ and $L_z$, can be approximated via
equations (\ref{orbitmaxs},\ref{sgen}) with 
\begin{equation}
\label{newv}
\Omega(i) = \frac{C(i)}{1-C(i)} \sum_{j \in {\cal J}(i), j\ne i} w(j),
\end{equation}
\begin{equation}
C(i) = V(i) \left( \sum_{j \in {\cal J}(i)} V(j) \right)^{-1}
\end{equation}
and
\begin{equation}
{\cal J}(i) = \left \{ j \in \{1,\ldots,N\} : L_z(j) = L_z(i), \ E(j) = E(i)\right \}
\end{equation}
($N$ is the total number of orbits). Equation (\ref{newv}) derives from the 
constraint that for $\felz \approx f(E,L_z)$, 
the phase-space density of any orbit $i$
with energy $E(i)$ and angular momentum $L_z(i)$ has to equal the mean phase-space
density of all orbits $j$ with the same $E(i)=E(j)$ and $L_z(i)=L_z(j)$, i.e.
\begin{equation}
f(i) = \frac{w(i)}{V(i)} \approx \frac{\sum_{j \in {\cal J}(i)} w(j)}{\sum_{j \in {\cal J}(i)} V(j)}.
\end{equation}

This case is included in Fig.~\ref{aniso_prof} by the circles.
That the $\Omega(i)$ from equation (\ref{newv}) indeed yield
$f \approx f(E,L_z)$ is demonstrated by $\beta \approx 0$. The flattening of 
the corresponding systems comes from an excess energy
in $\phi$-direction with respect to the isotropic case ($\gamma < 0$; orbits with
high angular momentum are strongly populated). The relationship
between $\delta$ and $\epsilon$ is similar as in maximum entropy models.

Note that DFs $f \approx f(E,L_z)$ develop noticeable phase-space density peaks
on orbits with high angular momentum \citep{DG94}.
It is likely this property that lowers their entropy as compared to the $\fmax$ models.
Flattening by anisotropy mainly involves shell orbits which approach closely the intrinsic
minor-axis. Their phase-space volumes are much larger than those of equatorial 
near-circular orbits with high angular momentum. Even a small change in the phase-space 
density along shell orbits can reduce the amount of light near the minor-axis considerably and, thus, result
in a significant flattening. The larger fraction of phase-space involved in this type of
flattening, compared with a strong overpopulation of the
relatively small region in phase-space occupied by near-circular orbits (as in cases
where $f = f(E,L_z)$) explains why objects which are flattened by anisotropy have the 
higher entropy.

\subparagraph*{Flattening with radial anisotropy.}
Model DFs $\frad$ obtained with
\begin{equation}
\Omega(i) = [r_\mathrm{apo}(i) - r_\mathrm{peri}(i)]^4 \times V(i)
\end{equation}
are biased towards orbits with a large difference 
$r_\mathrm{apo} - r_\mathrm{peri}$ 
between apocentre and pericentre radius (radially extended orbits). 
Such models are radially anisotropic ($\beta,\gamma>0$; cf. triangles in 
Fig.~\ref{aniso_prof}). The relationship between
$\delta$ and $\epsilon$ is again similar as in the previous models.

The latter is no surprise, as for self-consistent ellipsoids with constant flattening,
$\delta(\epsilon)$ can be calculated from the tensor virial theorem \citep{Rob62,Bin87}:
\begin{equation}
\label{eq:deltavir}
\delta(\epsilon) = 1 - \frac{1}{q(e)},
\end{equation}
where
\begin{equation}
q(e) = \frac{0.5}{1-e^2} \times \frac{\arcsin(e)-e\sqrt{1-e^2}}{e(1-e^2)^{-0.5}-\arcsin(e)}
\end{equation}
and
\begin{equation}
e = \sqrt{1-(1-\epsilon)^2}.
\end{equation}
The solid line in the upper-left panel of Fig.~\ref{aniso_prof} shows
relation (\ref{eq:deltavir}). Our numerically constructed orbit models follow this line well.

Note that, if $N$ DFs $f_i$ project each to the same spatial density, then any 
convex linear combination $f_\lambda = \sum \lambda_i \, f_i$ with $\sum \lambda_i = 1$ 
will do so. The properties of $f_\lambda$ will be intermediate 
between those of the individual $f_i$.

%
\begin{figure*}\centering
\begin{minipage}{170mm}
\centering
\includegraphics[width=169mm,angle=0]{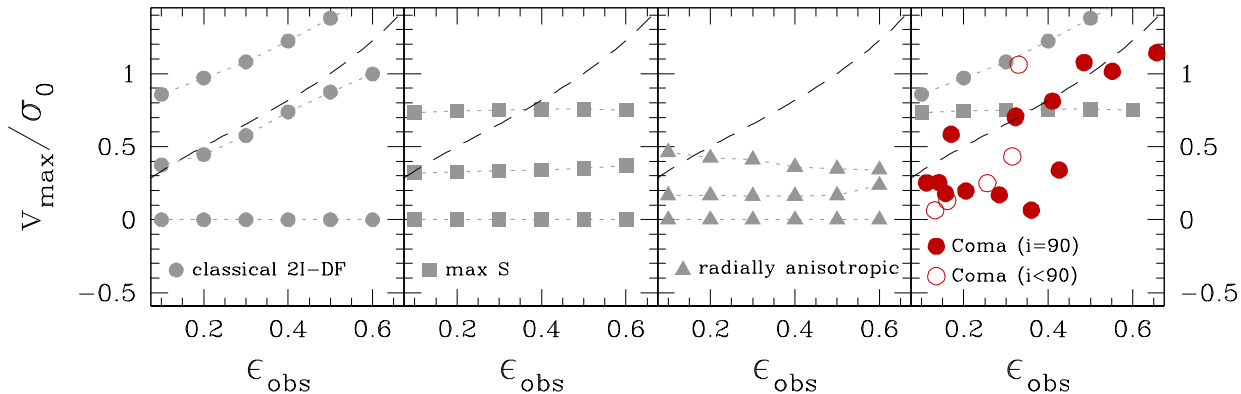}
\caption{First three panels, from left to right: ellipticity 
$\epsilon_\mathrm{obs}$ versus 
classical $v/\sigma$ for the toy models $\felz$ (grey circles), $\fmax$ (grey squares) and 
$\frad$ (grey triangles). For each toy model the cases $\mu = 0.5,0.7,1.0$ 
(no rotation, intermediate and maximum rotation) are shown. Models with the same $\mu$ are
connected by the dotted lines. All toy models are
edge-on such that $\epsilon_\mathrm{obs} = \epsilon$. Outer right panel: 
$\epsilon_\mathrm{obs}$ versus $v/\sigma$ for Coma galaxies (filled circles: edge-on galaxies,
open circles: non edge-on galaxies). For comparison, the maximally rotating $\felz$ and 
$\fmax$ models are also shown in the outer right panel. In all panels, the dashed lines 
approximate edge-on isotropic rotators (cf. equation \ref{eq:isorotapp}).}
\label{eps_vsig}
\end{minipage}
\end{figure*}

\subsection{Rotation}
\label{subsec:rot}
The just discussed toy models (and any linear combination of them) 
are non-rotating, because in our choices for $\Omega(i)$ 
we haven't distinguished between prograde and retrograde orbits. A large variety of rotation 
patterns can be constructed from any DF $f$ as follows: each orbit in an axisymmetric potential comes in two
flavours, one prograde (with positive $L_z>0$) and one retrograde ($L_z<0$). Both
share the same spatial shape but differ only in the sign of the velocity
component around the axis of symmetry. Thus, the spatial density will only depend on the 
sum
\begin{equation}
\shapefunc(E,L_z,I_3) \equiv f(E,L_z,I_3) + f(E,-L_z,I_3),
\end{equation}
of light on corresponding prograde and retrograde orbits. The amount of rotation, instead, 
will depend on the difference between the population of the prograde and retrograde
orbits, respectively. This can be quantified, for example, by the fraction $\mu$ of light 
on the prograde of each orbit pair:
\begin{equation}
\mu(E,L_z,I_3) = \frac{f(E,L_z,I_3)}{\shapefunc(E,L_z,I_3)}
\end{equation}
($L_z\ge0$).
For simplicity, let's assume from now on that $\mu$ is the same for all orbits. 
Then, any
\begin{equation}
\label{flammu}
f_{\mu}(E,L_z,I_3)\equiv\left\{ 
\begin{array}{l@{\quad:\quad}l}
\mu \, \shapefunc(E,L_z,I_3) & L_z \ge 0\\
(1-\mu) \, \shapefunc(E,-L_z,I_3) & L_z < 0
\end{array}
\right.
\end{equation}
with $\mu \in [0,1]$ (to remain positive definite)
will give rise to the same density profile as $f$ ($\shapefunc_\mu \equiv \shapefunc$), but with 
different degrees of internal rotation. For example, in case of $\mu \equiv 1/2$ prograde
and retrograde orbits are populated equally and there will be no rotation in the
corresponding system. With $\mu \equiv 1$ ($\mu \equiv 0$) only prograde (retrograde) orbits
are populated (maximum rotation).

The bottom panel of Fig.~\ref{aniso_prof} shows anisotropies for the toy models
of Sec.~\ref{subsec:toy} with $\mu = 0.9$. 
While $\beta$ is independent of the amount of rotation, $\delta$
decreases and $\gamma$ increases with increasing rotation. 
The latter reflects that in our toy models the total energy in $\phi$-direction is constant.
Any increase of the ordered motion is thus to the expense of a smaller $\sigma_\phi$.

Fig.~\ref{eps_vsig} illustrates where the toy models appear in the $(v/\sigma,\epsilon)$ 
diagram. The figure shows the three cases $\mu = 0.5$ (no rotation), $\mu = 0.7$
(intermediate rotation) and $\mu = 1.0$ (maximum rotation). On the y-axis, the ratio 
$v_\mathrm{max}/\sigma_0$ of the maximum rotation velocity ($v_\mathrm{max}$, along the 
projected major-axis) and the central velocity dispersion ($\sigma_0$, averaged
inside $\reff/2$) is shown. All models are edge-on. The highest rotation
rates at a given flattening are obtained for $\felz$, because of its strongly populated
high angular momentum orbits ($\Pi_{\phi\phi} > \Pi_{RR} \approx \Pi_{zz}$). 
However, $\fmax$ models, which are not flattened by an excess of light on high angular 
momentum orbits (relative to the isotropic case) but instead by a suppression of orbits with 
large $z$-velocities ($\Pi_{zz} < \Pi_{RR} \approx \Pi_{\phi\phi}$)
can reach $(v_\mathrm{max}/\sigma_0) \approx 0.75$ as well.
The dashed lines in Fig.~\ref{eps_vsig} approximate classical 
isotropic rotators by
\begin{equation}
\label{eq:isorotapp}
\left( 
\frac{v_\mathrm{max}}{\sigma_0}
\right)_\mathrm{iso} \equiv \sqrt{ \frac{\epsilon}{1-\epsilon}}
\end{equation}
\citep{Kor82}. Up to $\epsilon \approx 0.4$, $\fmax$ models can
appear in the same region as classical isotropic rotators, although they are not
flattened by rotation in the classical sense (e.g. $\beta \ne 0$). Radially anisotropic
models are dominated by orbits with low angular momentum and have generally low rotation
rates. 

%
\begin{figure*}\centering
\begin{minipage}{170mm}
\centering
\includegraphics[width=165mm,angle=0]{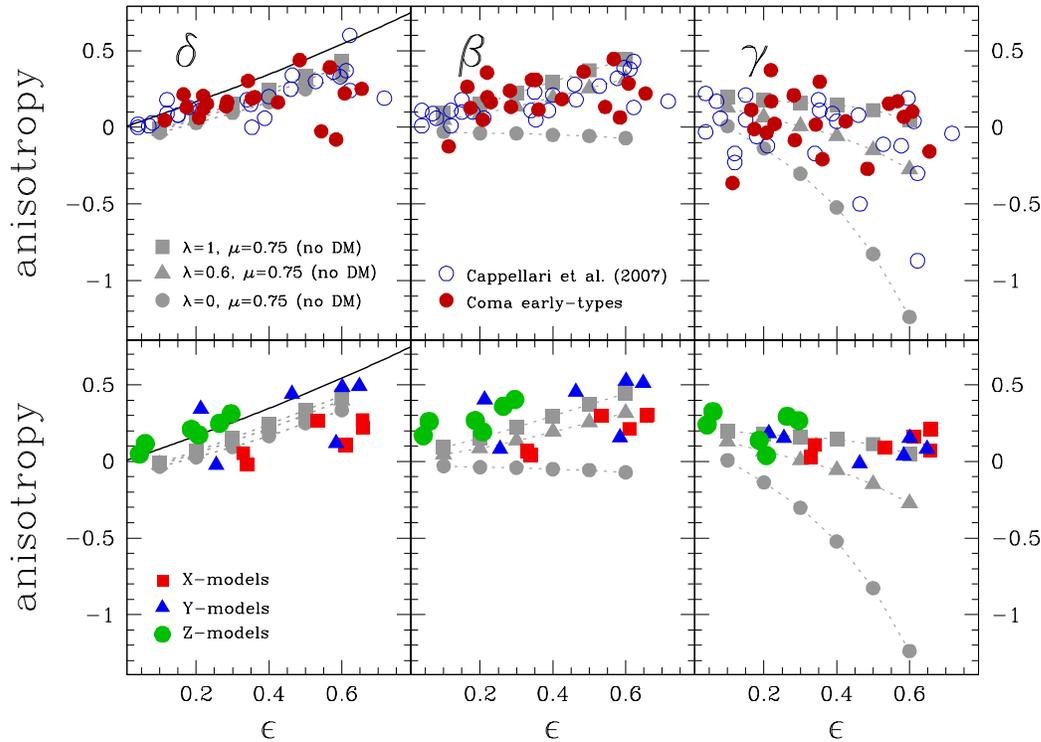}
\caption{Top: Anisotropy parameters $\delta$, $\beta$ and $\gamma$ (as labelled 
in the panels) versus intrinsic ellipticity $\epsilon$ for real galaxies (cf. mid panel
in top row). Bottom: same for models of $N$-body merger remnants. Light: toy models
from equation (\ref{eq:toy}), connected by dotted lines. Solid lines in left panels: tensor virial theorem applied to
oblate spheroids with constant flattening.}
\label{aniso_eps}
\end{minipage}
\end{figure*}

A complete picture of an axisymmetric 
galaxy's flattening mechanism requires knowledge of the 
amount of rotation (e.g. $v/\sigma$) and at least one anisotropy parameter 
(e.g. $\beta$, $\gamma$ or $\delta$ or the parameter $\alpha$ in the notation
of \citealt{Bin05}). Alternatively, two anisotropy parameters also specify the
global orbital structure. In any case, the full information about the anisotropy
and the flattening mechanism cannot be provided by the ($v/\sigma,\epsilon$) diagram 
alone.
For example, four among the five intrinsically most flattened \coma early-types are very close to the
isotropic rotator line in Fig.~\ref{eps_vsig}. However, they are shaped by a combination 
of $\beta \ga 0$ and $\gamma \la 0$ (cf. Tab.~\ref{tab:aniso}).

\subsection{Influence of a dark matter halo}
The presence of dark matter around
a galaxy affects the shape of the stellar orbits. Some of the
models just discussed may not exist, if an additional dark matter halo reshapes the
potential significantly. To check this, we have recalculated all our toy models in
a potential, where a spherical, logarithmic dark halo has been added to the
stellar potential. The parameters of the halo (its core radius and its asymptotic circular
velocity) have been set according to the dark matter scaling relations in \coma 
early-types \citep{Tho08}. The derived anisotropies in the new potential 
differ in no case by more than $0.1$ from the original ones (but see the discussion in 
App.~\ref{app:entropy}). Especially, the relationship between
$\beta$ and $\epsilon$, that arises from the maximisation of the orbital entropy also
appears in potentials with a realistic dark matter halo. This 
does not necessarily imply that 
the neglect of dark matter in models of real galaxies 
has no effect on the derived anisotropies, because it may enforce a redistribution
of the orbits (cf. next Sec.~\ref{sec:galaxies}).

\section{Real galaxies}
\label{sec:galaxies}
Fig.~\ref{aniso_eps} shows the connection between anisotropy and flattening
in real galaxies. The intrinsic flattening of \coma galaxies is expressed in terms of 
\begin{equation}
\label{eq:eps}
\epsilon = \frac{\int \der R \, R^2 \, \mathrm{SB}(R) \, \epsilon(R)}
{\int \der R \, R^2 \, \mathrm{SB}(R)}
\end{equation}
\citep{Bin05}. Here, $R$ is the radius along the 
projected major-axis and $\mathrm{SB}(R)$ and $\epsilon(R)$ are the 
surface-brightness profile and ellipticity profile in the edge-on projection. For an axisymmetric
system (with flattening q) $\epsilon \approx (1-q)$. 

Lines in Fig.~\ref{aniso_eps} trace three different toy models 
\begin{equation}
\label{eq:toy}
f_\lambda = \lambda \, \fmax + (1-\lambda) \, \felz
\end{equation}
(cf. Sec.~\ref{sec:theory}; the three models are designed to rotate by 
using $\mu = 0.75$ in equation \ref{flammu}). DFs $f \approx f(E,L_z)$ 
are inconsistent with the global orbital structure of most galaxies 
(because $\beta >0$ in observed galaxies). Most galaxies have orbital properties
between those of $\fmax$ and $\felz$ (with some rotation).

Fig.~\ref{aniso_eps} also includes anisotropies and flattenings of 24
early-types from \citet{Cap07}. These galaxies
are a subsample of the 48 Es/S0s of the SAURON survey \citep{deZ02}, which
uniformly covers the plane of observed flattening $\epsilon_\mathrm{obs}$ and $M_B$ (for
$M_B < -18$). The galaxies of \citet{Cap07} are
drawn from this survey according to various requirements, among them
consistency with axial symmetry (according to 2d kinematical maps). The galaxies
of \citet{Cap07} (shortly \sau in the following) are on average fainter than the
\coma galaxies. 

Although the samples do not match exactly, 
the anisotropies of \coma and \sau galaxies are found in the same
range. However, the \coma sample contains relatively more anisotropic but nearly round 
galaxies on the one hand and more highly flattened but isotropic galaxies ($\delta \approx 0$) 
on the other. As a result, the trend for $\delta$ and $\beta$ to increase with
$\epsilon$ which is seen in the \sau sample is not obvious when considering the complete 
\coma sample (even not if the two Coma galaxies with the most uncertain anisotropies 
are ignored -- the two central galaxies GMP2921 and GMP3329). 

The relation between $\beta$ and $\epsilon$ is weaker in the \coma galaxies 
in part due to a few round but anisotropic galaxies -- for example GMP1750 and GMP5568 with 
$\epsilon \approx 0.2$ and $\beta \approx 0.26-0.36$.  Both galaxies show weak minor-axis 
rotation \citep{Tho07} and could be slightly triaxial systems. In addition to differences
among nearly round galaxies, anisotropies in \coma and \sau galaxies also slightly
differ at high $\epsilon$. The latter is most clearly seen in $\delta$ versus
$\epsilon$: two highly flattened \coma galaxies 
(GMP1990 and GMP2440, $\epsilon \approx 0.6$) have $\delta \approx 0$. 
One of these galaxies is likely close to edge-on, because of its high
observed ellipticity ($\mathrm{max} \, \epsilon_\mathrm{obs} \approx 0.625$, cf. the radial
profile in \citealt{Tho07}) and its significant isophotal shape distortions. 
We expect the model of GMP1990 to be well constrained, because of the near edge-on 
inclination (minimal uncertainties in the deprojection) and its far-extending 
multi-slit kinematic data. 
For the other galaxy (GMP2440) \citet{Meh00} quote only a modest observed
ellipticity $\epsilon_\mathrm{obs} = 0.33$ at $\reff$ and the intrinsic flattening comes mostly
from the low inclination of the model. Note that this
galaxy is far above the isotropic rotator line in the right panel
of Fig.~\ref{eps_vsig} (GMP2440 is the only non edge-on galaxy above the
isotropic rotator line). A maximum-entropy like DF is ruled out for this
galaxy, because even the maximally rotating version of the $\fmax$ model would not allow
for the high observed rotation rate. Thus, even if we would have underestimated the inclination
of this system, Fig.~\ref{eps_vsig} shows that its orbital structure must be significantly 
deviant from maximum-entropy models. All in all then, modelling uncertainties are unlikely to explain the outstanding 
anisotropies of GMP1990 and GMP2440. In fact, a comparison 
with Fig.~3 in \citet{Cap07} reveals that the \sau sample does not include galaxies like 
GMP1990 and GMP2440, because (1) for only one object the
observed ellipticity is significantly larger than $\epsilon_\mathrm{obs} > 0.5$ (NGC4550) and
(2) even the fastest rotators in the \sau sample are closer to the isotropic rotator line
than GMP2440. 

In addition to differences in the sample selection also the modelling methods
differ in the details. \citet{Cap07} use similar orbit-based dynamical models as 
we do here, but \sau anisotropies are calculated
inside a fixed aperture
with a radius of $25\arcsec$. A fixed aperture encloses different fractions of the 
stellar mass in different galaxies, depending on system 
size and distance. For the \coma galaxies we give anisotropies inside $\reff$. 
In some galaxies local anisotropies
vary significantly with radius \citep{Tho07}, such that the radius of comparison is crucial.
In addition, \sau models are based on the assumption that mass follows light.
As it has been stated already 
in the introduction, the assumption of a constant mass-to-light ratio can result in 
artificially large $\phi$-energies
(\citealt{Tho05}, \citealt{For08}) or low $\gamma$, respectively. Regarding
Fig.~\ref{aniso_eps}, \sau galaxies do not have systematically lower $\gamma$ than \coma
objects. For the only two exceptions (NGC4473 and NGC4550) \citet{Cap07} 
report evidence for counter-rotating, disk-like components that likely cause
their large $\phi$-energies. 
The small effect that the neglect of dark matter has on the anisotropies
likely reflects the fact that we only consider anisotropies
averaged inside $r \la \reff$, where the assumption that mass
follows light is most closely fulfilled (e.g. \citealt{G01}, \citealt{Tho07}).
For the Coma galaxies a quantitative comparison of models with and without dark matter
is made in App.~\ref{sec:app}.

The spatial coverage with kinematic data in the inner regions is sparse in the
\coma galaxies (long-slit data) compared to the \sau objects (2d kinematical
maps). In regions of phase-space that are not well constrained by the observed kinematics,
the dynamical models are mainly driven by regularisation. Thus, because the spatial coverage
is lower in the \coma galaxies, their anisotropies could be biased towards the 
adopted regularisation scheme. Specifically, \coma galaxy models are regularised 
towards maximum entropy \citep{Tho05}. However, the middle panel in the top row of 
Fig.~\ref{aniso_eps} does not show any bias of the \coma models towards the maximum
entropy relation. In fact, \sau galaxies are on average closer to this
relation than \coma galaxies. This indicates that regularisation is not the main driver 
for the \coma galaxy models. Also, in App.~\ref{sec:app} we give an explicit comparison 
of \coma galaxy models with standard and with weak regularisation.
We do not find significant differences. 

Both the intrinsic ellipticity and the anisotropy depend on the inclination
of the models. For the \coma galaxies, we probe three different inclinations and 
use the one that fits best \citep{Tho07}, while inclinations for \sau galaxies
are derived from two-integral Jeans models \citep{Cap06}. 
The inclination is best constrained for highly flattened galaxies, because
these have to be close to edge-on. For three of the \coma galaxies 
(GMP0756, GMP1176 and GMP1990) large
ellipticities together with significantly discy/boxy isophotes indeed indicate 
close to edge-on inclinations (for example, GMP1176 exhibits $a_4 > 10$; 
\citealt{Cor08}). In contrast, two among the five galaxies
with $\epsilon \ge 0.5$ owe their flattening in part from the relatively low
inclination of the best-fit model (GMP0282, GMP2440; cf. Tab.~\ref{tab:aniso}). 
These galaxies provide the smallest and largest anisotropies, respectively, 
at high $\epsilon$ (cf. middle panel in the top row of Fig.~\ref{aniso_eps}).
This suggests that the method to determine the inclination for the \coma galaxies
does not result in a specific bias of the derived anisotropies.

We conclude that slight differences between the \sau and the \coma anisotropies are mostly
due to the different sample selections, while differences in the modelling
methodology (including differences in the data coverage) seem to be negligible.

%
\begin{figure*}\centering
\begin{minipage}{170mm}
\includegraphics[width=165mm,angle=0]{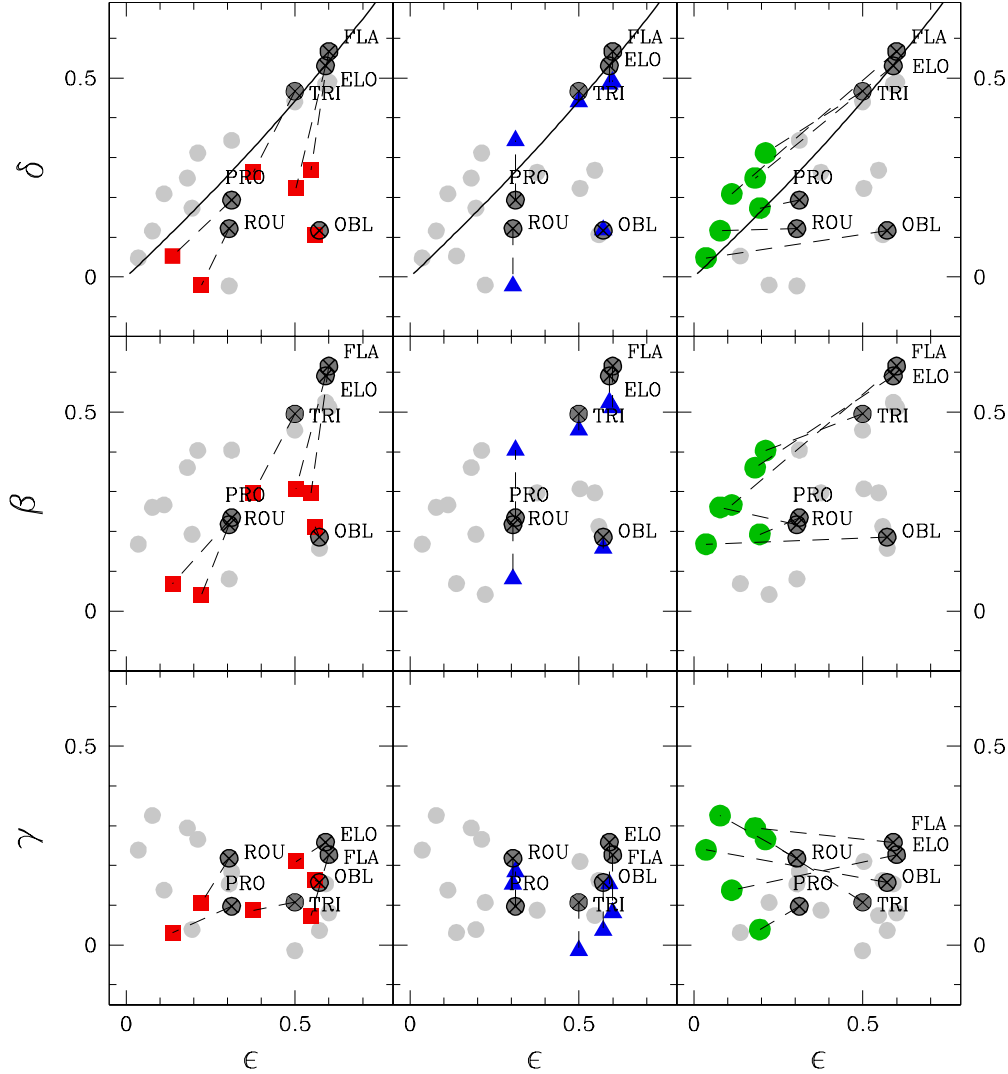}
\caption{From top to bottom: anisotropy parameters $\delta$, $\beta$ and $\gamma$ 
versus (intrinsic) ellipticity $\epsilon$. Heavy symbols: intrinsic parameters of
$N$-body merger remnants (abbreviations refer to the TRIAX, PROLATE, ROUND, FLAT, ELONG and 
OBLATE merger remnants discussed in \citealt{Tho07b}); 
light: Schwarzschild models of merger remnant projections. From left to right:
models of $X$, $Y$ and $Z$-projections are highlighted in colour. In each panel, 
dashed lines indicate which model belongs to which merger remnant. Solid lines in top row:
$\delta(\epsilon)$ of equation (\ref{eq:deltavir}), as in Fig.~\ref{aniso_prof}.}
\label{aniso_recon}
\end{minipage}
\end{figure*}

\section{Comparison of observed galaxies and $N$-body merger remnants}
\label{sec:merger}
The lower panels of Fig.~\ref{aniso_eps} display the models of
$N$-body merger remnants (cf. Sec.~\ref{sec:data}). In terms of $\delta$ versus $\epsilon$ and
$\beta$ versus $\epsilon$, these models do not differ strongly from 
models of real galaxies (see also \citealt{Bur07}). However, while $\gamma \ge 0$ 
in models of merger remnants, $\gamma$ is often negative in models of real galaxies. 
Is this discrepancy in $\gamma$ indicative for the merger remnants having a different
orbital structure than real galaxies, or 
does it merely reflect systematics caused by the symmetry assumptions 
in our models?

Reconstructed and true intrinsic anisotropies\footnote{Note that while $\pix = \piy$ in
axisymmetric systems, $\pix \ge \piy$ in the merger remnants. For the intrinsic
$\delta$ of the merger remnants we use the average $(\pix + \piy)/2$ instead
of $\pix$ in equation (\ref{eq:delta}).} and flattenings of the merger 
remnants are compared in Fig.~\ref{aniso_recon}. The one merger remnant closest to oblate axial symmetry 
(OBLATE), is reconstructed with high accuracy from the X and Y-projections (edge-on). 
This is plausible, because for this remnant the assumption of axial symmetry 
is a good approximation. Furthermore, in the edge-on case the deprojection becomes unique and the
intrinsic degeneracies in the dynamics are likely smallest. 

However, the general trend in the axisymmetric models is
to underestimate both, the flattening and the anisotropy of the merger remnants. 
X and Y-projections allow a better reconstruction of shape and anisotropy
than Z-projections. It has already been discussed in \citet{Tho07b} that 
the assumption of axial symmetry enforces an inclination mismatch in the Z-models: 
while the triaxial remnants appear flattened in the $Z$ projection (face-on),
axisymmetric systems are necessarily round when seen face-on. Then, because the
models are forced towards a wrong viewing-angle, (1) the intrinsic flattening is
underestimated and (2) X, Y and Z-axes of models and
remnants do no longer correspond to each other. For example, a Z-model's $\gamma$ 
measures a different energy ratio as the remnant's $\gamma$ \citep{Tho07b}. Had we compared
the Z-models with the apparent shape of the remnant in Z-projection and with the 
energy ratios along axes of models and remnants that correspond to each other, then the
differences would have been much smaller (for example 
$\Delta \epsilon < \Delta \delta \la 0.1$).

%
\begin{figure*}\centering
\begin{minipage}{170mm}
\centering
\includegraphics[width=125mm,angle=0]{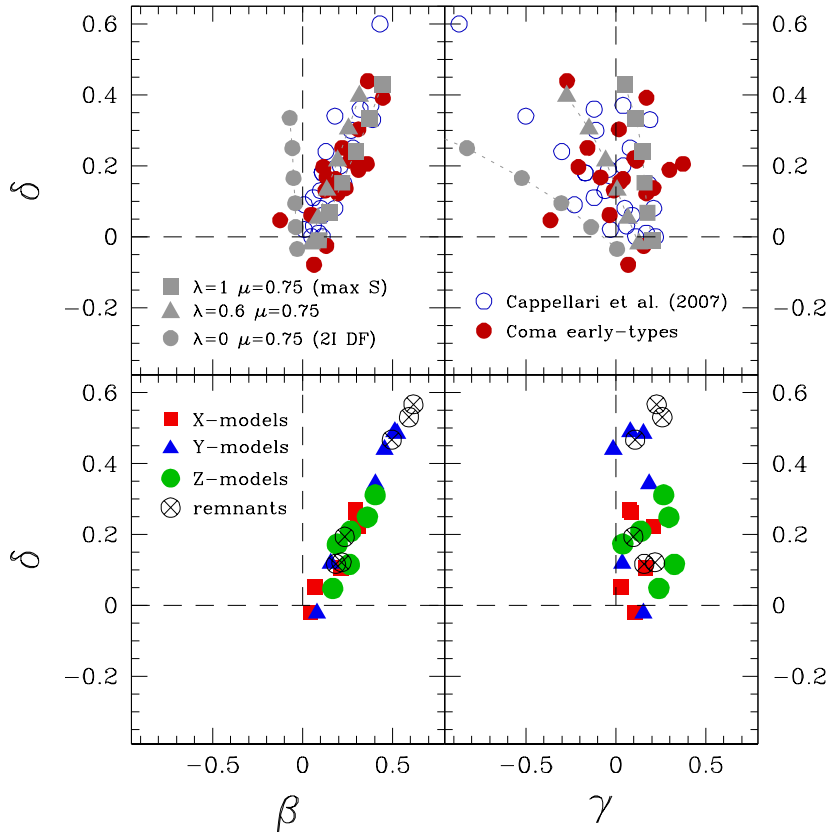}
\caption{Correlations among the anisotropy parameters $\delta$, $\beta$ and 
$\gamma$. Top: models of real galaxies (squares, triangles and circles: toy models
from equation \ref{eq:toy}, connected by dotted lines); bottom: models of merger remnants (squares,
triangles and circles separate models of projections along different 
principal axes as indicated in the lower-left panel; crosses with circles: intrinsic 
anisotropies of the $N$-body remnants).}
\label{aniso_aniso}
\end{minipage}
\end{figure*}

Real galaxies are seen at random viewing angles. Starting from our
models of principal projections it is difficult to
predict directly the analogous distributions of $\delta(\epsilon)$ or $\beta(\epsilon)$ 
for the realistic case of random projections.
However, because the projections along principal axes yield extreme kinematical
and photometrical properties of the merger remnants \citep{Jes05}, it can be expected
that dynamical models of 
projections along intermediate viewing-angles will have properties intermediate
between those of the models from principal projections. We have verified this for two out 
of the six merger remnants (ELONG and OBLATE) by modelling additional 11 projections (at 
intermediate viewing-angles). Assuming that this result can be generalised to
other remnants as well, then
Figs.~\ref{aniso_eps} and \ref{aniso_recon} suggest the following: if
real galaxies would resemble the modelled merger remnants, then one would see 
approximately the same relationships $\delta(\epsilon)$ and $\beta(\epsilon)$ as 
in the \coma and \sau galaxies. However, it is
clear from Fig.~\ref{aniso_recon} that $\gamma \ge 0$, for a sample of randomly projected
objects like our modelled $N$-body merger remnants. Thus, in this respect, models of 
many real galaxies differ from our comparison sample of synthetic
$N$-body merger remnants: models of merger
remnants are always radially anisotropic ($\Pi_{RR} > \Pi_{\phi\phi} \approx \Pi_{zz}$), 
while models of real galaxies are characterised by $\Pi_{\phi\phi} \ga \Pi_{RR} > \Pi_{zz}$.

This fact is further illustrated in Fig.~\ref{aniso_aniso}, which 
shows correlations among the anisotropy parameters. Even though shape and anisotropy 
cannot be recovered simultaneously (in some cases),
the anisotropy correlations in the models of the merger remnants and in the
merger remnants itself are very similar to each other. Again, the main difference
between real galaxies and merger models is the offset between their $\gamma$ distributions.

Besides the fact that merger remnants have on average positive $\gamma >0$, while real
galaxies have $\gamma \approx 0$ (on average), Fig.~\ref{aniso_aniso} shows that the
distribution of anisotropies in the merger remnants is tighter than in real galaxies.
This may reflect the similarity in the initial conditions of the $N$-body simulations
(most noteworthy the similarity in the progenitors and the fact that we only consider
collisionless mergers).

\section{Implications for the formation process of early-type galaxies}
\label{sec:discussion}
The anisotropy parameters defined in 
Sec.~\ref{sec:theory} are only global measures of the orbital structure. A full
understanding of the formation process of early-type galaxies can only be provided
by spatially resolved anisotropy profiles. For example, equatorial near circular
orbits obey, in the epicycle approximation, the local relation
\begin{equation}
\label{eq:epicyc}
\frac{\sigma_\phi^2}{\sigma_R^2} \approx \frac{1}{2} 
\left( 
1 + \frac{\der \ln v_\mathrm{circ}}{\der \ln r}
\right),
\end{equation}
where $v_\mathrm{circ}$ is the circular velocity \citep{Bin87}. 
In a typical galaxy potential
the circular velocity curve is flat ($\der \ln v_\mathrm{circ} / \der \ln r \approx 0$) and
equation (\ref{eq:epicyc}) predicts 
$\sigma_\phi^2 \approx 0.5 \, \sigma_R^2$. Since the epicycle approximation holds for
perturbed rotating disks, we do not expect the majority of early-types in our sample to be well
described by equation (\ref{eq:epicyc}). However, it might be
relevant for the most flattened, rotating and discy objects in our sample.
Instead, at least some of these (for example GMP1176 and GMP3958) have 
negative $\gamma$ (i.e. $\sigma_\phi > \sigma_R$). This does not rule out a 
disk heating scenario for these galaxies, however, because locally we find 
$\sigma_\phi^2 \approx (0.5 \cdots 0.7) \, \sigma_R^2$ 
near the equatorial plane in these galaxies (around $r \approx \reff$; cf. the 
radial anisotropy profiles in \citealt{Tho07}). 

In case of the collisionless $N$-body merger simulations, already the
averaged anisotropy parameters reveal significant differences to the models of
real galaxies. Which physical processes are responsible for this discrepancy?

The orbital structure of the models of merger remnants is
largely driven by a population of central box orbits in the $N$-body systems \citep{Tho07b}.
They cause the centres of the merger remnants to become triaxial/prolate and are, for example,
largely responsible for the wrong viewing angle of the Z-models.
Dissipation during a merger can have a significant effect on the shape and the projected 
properties of the final remnant \citep{Bar96,Cox06,Rob06,Nab06,Jes07}. Already 10 percent of gas are sufficient 
to suppress central box orbits and to produce an approximately axisymmetric remnant 
in binary mergers \citep{Nab06} (but this result is based on simulations without
star formation). 

Multiple, simultaneous minor mergers likewise produce remnants less triaxial than collisionless
binary merger remnants \citep{Wei96}, but the corresponding kinematics have not yet been studied 
in detail. Successive minor merging does not necessarily lead to different final
remnants, at least if the cumulative merged mass becomes similar to the most 
massive progenitor \citep{Bou07}. Again, detailed predictions for the orbital
make-up and the shapes of the line-of-sight velocity distributions have not yet been
worked out.

Note that the central dark matter densities in \coma ellipticals are larger
than in present-day spirals \citep{Tho08}. Even if ellipticals have formed by some variant
of merging, present-day spiral galaxies are unlikely the progenitors for the bulk
of giant ellipticals (see also \citealt{Nab07a}). \citet{Bur07} 
pointed out that $N$-body systems, 
which have assembled hierarchically in their cosmological
simulations \citep{Nab07}, or by binary mergers with star-formation and black-hole feedback
are consistent with the trend between $\delta$ and $\epsilon$ in observed
galaxies.

\section{Summary}
\label{sec:summary}
We have discussed the relationship between anisotropy and flattening in toy models, 
in models of real galaxies, in merger remnants and in models of merger remnants. 
Models of observed galaxies generally exhibit $\beta > 0$ and $\gamma \approx 0$. 
We do not
find strong correlations of the anisotropy parameters $\delta$, $\beta$ and $\gamma$
with intrinsic ellipticity $\epsilon$.

In toy models with maximum entropy for a given density distribution we find
$\beta$ to increase with $\epsilon$, while $\gamma \la 0$. Observed galaxies appear
close to these maximum-entropy relations, but exhibit a large degree of individuality.
Rotation appears in anisotropic ($\beta > 0$) as well as isotropic systems ($\beta \approx 0$),
suggesting that the flattening of the galaxies largely
arises from a suppression of stars with large energies perpendicular to the equatorial
plane. This is similar to the classical notion of flattening by anisotropy and rules out 
DFs $f \approx f(E,L_z)$ for most early-type galaxies. 

The global similarity between models of observed galaxies and our maximum-entropy
toy models suggests that early-type galaxies are largely relaxed stellar systems. However,
there are differences in the details that probably contain valuable information about the
assembly mechanism of the galaxies and will be addressed in a future paper. 

Numerical simulations indicate that both strongly radially anisotropic
($\gamma \to 1$) and strongly tangentially anisotropic systems ($\gamma \ll 0$) can become 
unstable (e.g. \citealt{Mer90,Sel94,Nip02}). 
Maximum entropy models have intermediate anisotropies and are 
likely stable. Thus, the anisotropies
of observed galaxies may not only be understood as being the most likely 
ones (in the sense of yielding the maximum entropy at a given flattening) but could also
reflect stability constraints. So far we lack detailed studies exploring the
stability of axisymmetric systems with dark matter halos and 
various intrinsic anisotropies. Since our 
(three integral) toy models can be easily transformed to $N$-body systems (cf. \citealt{Tho07b}) they 
provide a suitable tool to setup both artificially anisotropic as well 
as realistic and observationally motivated initial conditions.

In models of real galaxies the unordered kinetic energy in the azimuthal direction,
$\Pi_{\phi\phi}$, can exceed the radial
energy $\Pi_{RR}$ by up to 40 percent. This separates 
real galaxy models from similar models of
collisionless $N$-body binary disk mergers, which are instead characterised
by radial anisotropy ($\Pi_{RR} > \Pi_{\phi\phi} \approx \Pi_{zz}$). Because we 
have applied the same modelling machinery to both, the real galaxies as well as the
synthetic $N$-body merger remnants, our findings indicate a true difference between their
intrinsic properties. Especially,
we have shown that if real galaxies would resemble our merger remnants, then 
corresponding dynamical models of real data would be radially anisotropic, irrespective of the
systematics introduced by the assumption of axial symmetry. 

The radial anisotropy of the merger remnants is related to a population
of central box orbits. Because dissipation during a merger can efficiently 
suppress box orbits, our results
suggest that dissipation played an important role during the formation of
intermediate mass to massive early-type galaxies.

In this paper we focussed on the comparison of real galaxies with collisionless 
binary disk merger simulations. A similar analysis, 
but for gaseous mergers with star formation and/or for 
galaxies formed in cosmological simulations could give
more insight into the actual formation paths of elliptical galaxies.

\section*{Acknowledgements}
This work was supported by DFG
Sonderforschungsbereich 375 ``Astro-Teilchenphysik''
and DFG priority program 1177. EMC receives support from grant
CPDA068415/06 by Padua University.

\appendix

\section{The influence of regularisation and dark matter on reconstructed galaxy anisotropies}
\label{sec:app}
%
\begin{figure*}\centering
\begin{minipage}{170mm}
\centering
\includegraphics[width=125mm,angle=0]{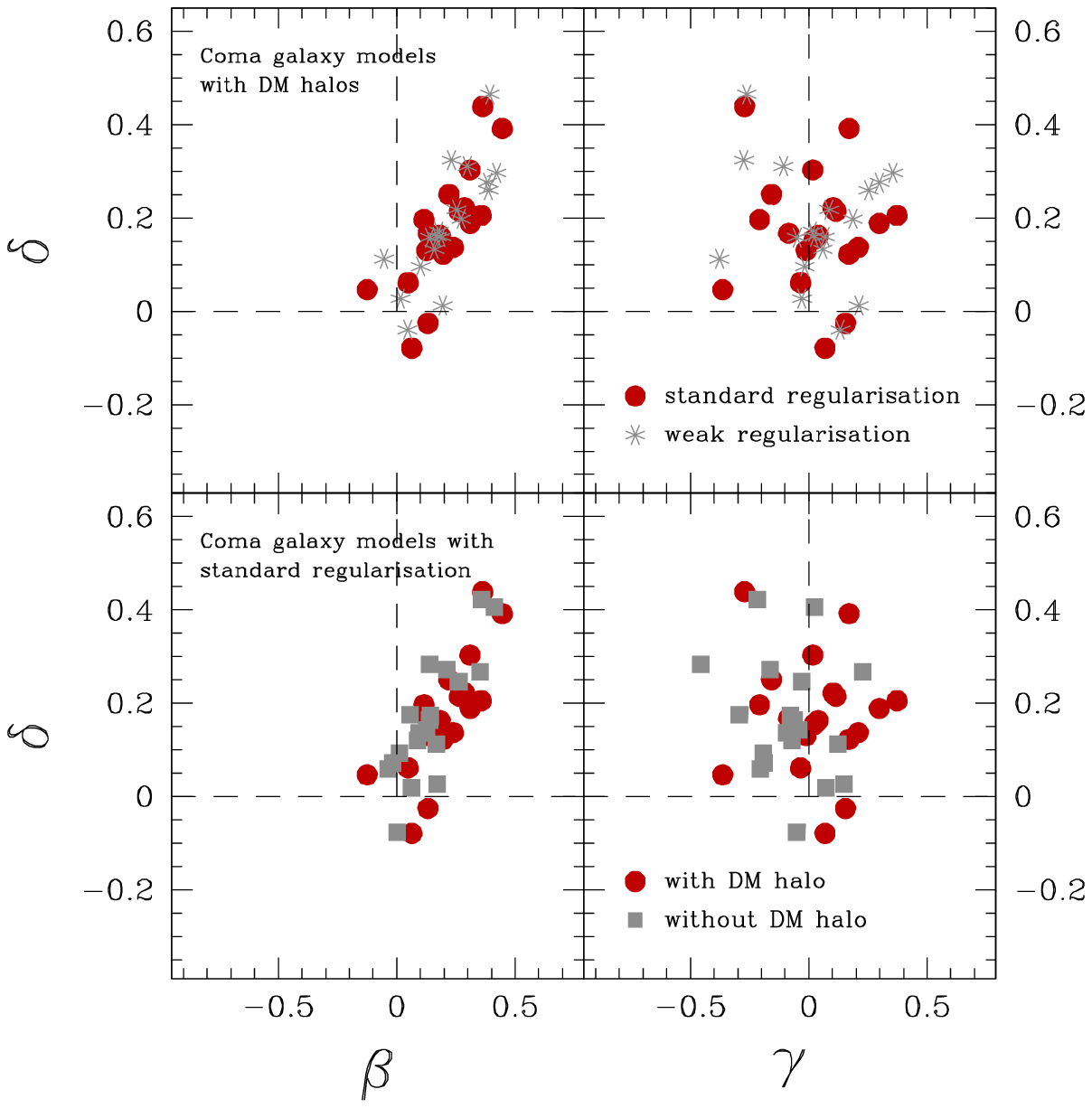}
\caption{As Fig.~\ref{aniso_aniso}, but in the top row Coma galaxy models are plotted 
with standard regularisation (filled circles; $\alpha =0.02$ in equation \ref{orbitmaxs}) and 
with weak regularisation (stars; $\alpha =1$). The bottom row shows Coma
galaxy models with (filled circles) and without (squares) dark matter halos (in both cases
the standard regularisation was applied).}
\label{aniso_aniso:test}
\end{minipage}
\end{figure*}
The Coma galaxy models discussed in this paper are derived using our standard 
regularisation parameter 
$\alpha = 0.02$ (cf. equation \ref{orbitmaxs}). The strength of the standard regularisation
has been calibrated by means of Monte-Carlo simulations of isotropic rotators \citep{Tho05}. To
check how much the choice of $\alpha$ affects our results, we (1) determined the
best-fit dynamical model at $\alpha =1$ and (2) recalculated the anisotropies of
all galaxies from these weakly regularised models (at $\alpha=1$, the
minimum $\chi^2$ is usually reached). In the top row of Fig.~\ref{aniso_aniso:test} we show
both models with standard and with weak regularisation for comparison. As can be seen, 
lowering the regularisation 
has almost no effect on the derived anisotropies. Especially, there are still at least four
galaxies with significantly negative $\gamma<0$.

In the bottom row of Fig.~\ref{aniso_aniso:test} we make a similar comparison
for models with and without dark matter: squares indicate the anisotropies of our
best-fit models with a constant mass-to-light ratio (no dark matter halo). 
As expected, when assuming that mass-follows-light, $\gamma$ become smaller (the amount
of $\phi$-energy is increased to compensate for the missing dark mass). 
From the bottom-right panel one would expect
that the average $\gamma$ becomes negative when the radial increase of the mass-to-light ratio
(caused by a dark halo) is neglected.
This is not the case in the \sau sample, however, although \citet{Cap07} assumed
that the mass-to-light ratio is constant with radius in their models. 
That neglecting dark matter
has a stronger effect in Coma galaxies is likely related to the fact that our kinematical
data reach out into the region where dark matter becomes noticeable 
($\ga \reff$), which is probably not the case in many \sau galaxies (where
the data extend only out to $\la \reff$).

\section{The radially resolved anisotropy structure of maximum-entropy toy models}
\label{app:entropy}
%
\begin{figure}\centering
\includegraphics[width=84mm,angle=0]{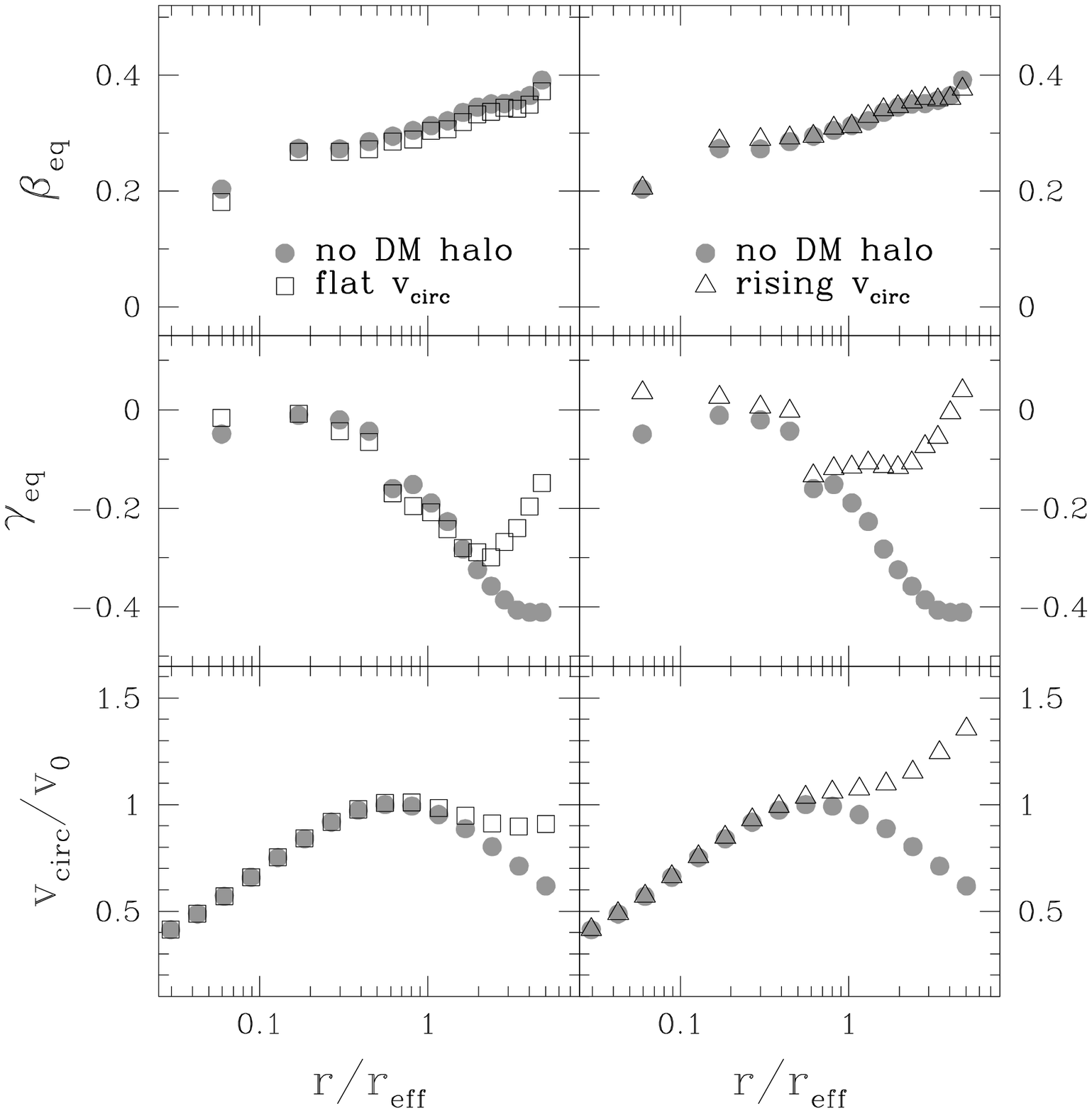}
\caption{Radial anisotropy profiles $\beta_\mathrm{eq}$ (top) and
$\gamma_\mathrm{eq}$ (middle) for maximum-entropy models ($\fmax$; flattening
$q=0.7$) with dark matter
halos. Left-hand side/squares: halo tuned to result in an approximately flat circular 
velocity curve; right-hand side/triangles: halo leading to an increasing circular velocity
in the outer parts of the model. For comparison, the
case without halo is shown on both sides (circles). Circular
velocity curves of the models (scaled to the maximum circular velocity $v_0$ without halo)
are shown in the bottom panels.}
\label{aniso_vcirc}
\end{figure}
The maximum-entropy toy models $\fmax$ discussed in Sec.~\ref{subsec:toy} resemble
the classical flattening by anisotropy, except
that they are only approximately isotropic in $R$ and $\Phi$. To investigate where
this anisotropy comes from, we have constructed maximum-entropy toy models in
potentials that include a dark matter halo. The halo density distribution is assumed
to follow
\begin{equation}
\label{eq:dehnen2}
\rho_\mathrm{DM} \propto m^{-\eta} (m+1)^{\eta-4}
\end{equation}
\citep{Deh93}. To mimick realistic halos (cf. \citealt{Tho07}) we choose a flat central density core 
($\eta = 0.05$) and we set the flattening $q$ of the halo equal to the
flattening of the luminous component of the toy model (cf. equation \ref{eq:flattening}).
We investigated three mass models: (1) no halo, (2) a mass model that has a
roughly constant circular velocity curve and (3) a mass model with a rising
$\vcirc$ in the outer parts of the model.
The corresponding circular velocity curves for $q=0.7$ are
shown in the bottom panels of Fig.~\ref{aniso_vcirc}.

%
\begin{figure}\centering
\includegraphics[width=84mm,angle=0]{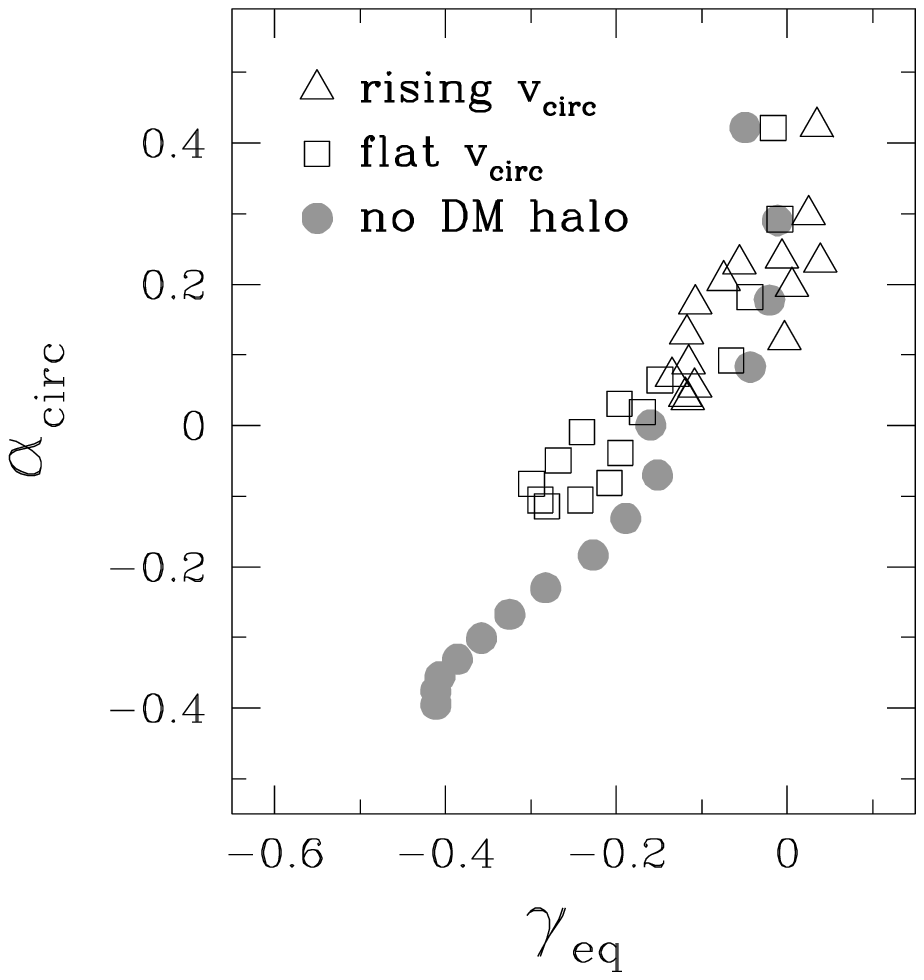}
\caption{Local anisotropy $\gamma_\mathrm{eq}$ (along the equatorial plane) versus
logarithmic slope $\alpha_\mathrm{circ} \equiv \der \ln v_\mathrm{circ} / \der \ln r$ 
of the circular velocity curve. Symbols and 
colours as in Fig.~\ref{aniso_vcirc}.}
\label{aniso_vcirc2}
\end{figure}

The upper panels of Fig.~\ref{aniso_vcirc} display the radial profiles of the local anisotropies
\begin{equation}
\beta_\mathrm{eq}(r) \equiv 1 - \frac{\sigma_z(r)^2}{\sigma_R(r)^2}
\end{equation}
and
\begin{equation}
\gamma_\mathrm{eq}(r) \equiv 1 - \frac{\sigma_\phi(r)^2}{\sigma_R(r)^2}
\end{equation}
along the equatorial plane (averaged within $|\vartheta| \le \pm 11.5\degr$, where $\vartheta$
is the latitude). As one can see,
the anisotropy in the meridional plane ($\beta_\mathrm{eq}$) does not depend
on the shape of the gravitational potential. Thus, the relation between $\beta$ and 
$\epsilon$ is largely independent 
from the gravitational potential and closely related to the entropy maximisation.

However, beyond $\reff$, where dark matter starts to influence the shape of the
circular velocity curve, $\gamma_\mathrm{eq}$ is different in the three different potentials.
Fig.~\ref{aniso_vcirc2} shows that the local
value of $\gamma_\mathrm{eq}$ -- along the equatorial plane -- is directly connected
to the logarithmic slope $\alpha_\mathrm{vcirc}$ of the circular velocity curve. 
In general then, because $\gamma$ from
equation (\ref{eq:gamma}) is the spatial average of $\gamma_\mathrm{eq}$ (and the
local anisotropies along other position angles in the meridional plane), its
exact value is not set uniquely by the entropy maximisation but also depends 
on the shape of the circular velocity curve. In practice, however, 
deviations with respect to the model without halo become noticeable only
beyond $\reff$, such that even the spatially averaged
$\gamma$ of the toy models does not depend strongly on whether a halo is included or not.

Note that the relation revealed by Fig.~\ref{aniso_vcirc2} is different from the
epicycle relation (\ref{eq:epicyc}). This is expected, because the azimuthal 
velocity dispersion
$\sigma_\phi$ in the toy models largely results from the fact that they do not
rotate. Instead, the dispersion predicted by the epicycle approximation arises
from perturbations on circular orbits in a rotating disk.

\label{lastpage}
\end{document}